\documentclass[pra,twocolumn,amsmath,amssymb,groupedaddress,superscriptaddress]{revtex4}
\usepackage{graphics}

\def\tr{\textmd{tr}}

\def\RR{{\bf R}}
\def\rr{{\bf r}}
\def\ss{{\bf s}}
\def\qq{{\bf q}}
\def\pp{{\bf p}}
\def\be{\begin{equation}}
\def\ee{\end{equation}}
\def\xy{\textmd{xy}}
\def\zz{\textmd{z}}
\def\ah{\textmd{ah}}
\def\qq{{\bf q}}
\def\kk{{\bf k}}
\def\bea{\begin{eqnarray}}
\def\eea{\end{eqnarray}}

\def\PP{{\bf P}}

\begin{document}

\title{Quantum Manipulation of Trapped Ions in Two Dimensional Coulomb Crystals.}

\author{D. \surname{Porras}}
\email{Diego.Porras@mpq.mpg.de} 
\author{J.~I. \surname{Cirac}}
\email{Ignacio.Cirac@mpq.mpg.de} 

\affiliation{Max-Planck-Institut f\"ur Quantenoptik, Hans-Kopfermann-Str. 1, Garching, D-85748, Germany.}

\begin{abstract}
We show that a large number of ions stored in a Penning trap, and forming a 2D Coulomb crystal, provides an almost ideal system for scalable quantum computation and quantum simulation. In particular, the coupling of the internal states to the motion of the ions transverse to the crystal plane, allows one to implement two qubit quantum gates. We analyze in detail the decoherence induced by anharmonic couplings with in--plane hot vibrational modes, and show that very high gate fidelities can be achieved with current experimental set--ups. 
\end{abstract}

\date{\today}

\maketitle

The search for a physical system where quantum computation is feasible is at the focus of an intense theoretical and experimental activity \cite{Cirac.Zoller.physics.today}. 
Ion traps are by now among the most promising candidates for a many--qubit quantum processor. 
In this system, qubits are stored in internal electronic states, and collective vibrational modes of the ions allow us to induce quantum gates between them \cite{ion.qc.theory}.
Following this idea, the building blocks for quantum computation have already been demonstrated in experiments with a few qubits \cite{ion.qc.experiment}.
Most of the current efforts to scale up the size of current ion
quantum processors, rely on the fabrication of arrays of microtraps
\cite{ion.microtraps}, in which a large number of ions can be stored and shuttled. 
Even though an astonishing progress has been achieved in this direction in the last years, the scalability of this system still demands technical advances in microfabrication and trap design \cite{Chuang}.

Penning traps provide us with an alternative trapping scheme, where a large number of ions ($10^4$ -- $10^6$) can be confined by a potential with approximate cylindrical symmetry  \cite{Bollinger}. 
Axial confinement is induced by a static electric field, whereas radial confinement is a result of the rotation of the ions under an axial magnetic field.
If the axial confinement is strong enough, ions arrange themselves in a triangular lattice on a single plane, which corresponds to a classical two dimensional (2D) Wigner crystal. 
The appeal of this system lies on the fact that ions are naturally ordered in a 2D regular array, without the need of individual micropotentials. 
Furthermore, ions are separated by distances of the order of tens of microns, such that they are individually addressable by optical means \cite{Mitchell}.
Thus, ions in Penning traps may appear as ideally suited for quantum computation and quantum simulation. However, this system has never been considered for this task \cite{Tombesi}.        
First, because the complicated vibrational level structure of the 2D crystal makes it difficult to apply here schemes that require resolution of single vibrational modes.
Beside that, typical schemes usually rely on the coupling of qubits to modes in directions parallel to the crystal. In current experiments with Penning traps, Doppler cooling of ions has reached temperatures of at most 1 mK, which implies occupation numbers of $10^2$ -- $10^3$ in the in--plane vibrational modes, so that, it seems not to be possible to use them for quantum operations.

\begin{figure}
\resizebox{\linewidth}{!}{
    \includegraphics{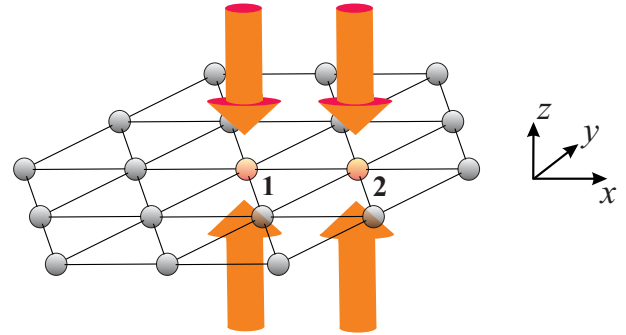}}
  \caption{(color online). Quantum gate in a 2D Coulomb crystal: standing--waves induce a state-dependent dipole force on two nearest neighbors in a triangular lattice.}
\label{axial.scheme}
\end{figure}

In this Letter we show how to circumvent these problems by exploiting the ions' motion along the {\em axial} direction (Fig. \ref{axial.scheme}). This approach benefits from the high axial confinement frequencies (and thus smaller occupation numbers at finite temperature), as well as from the fact that ions are weakly coupled in this direction, something that enormously simplifies the description of the ions' motion. In particular, we show how it is possible to carry out two-qubit gates between ions with high fidelities by performing a careful analysis of the main sources of decoherence. We emphasize that the results derived here also imply that this system is ideally suited for quantum simulations, which may be specially interesting due to the fact that ions are displayed in a triangular structure and favor the simulation of magnetic frustrated systems.

The main source of decoherence in our scheme is due to the anharmonic terms in the Coulomb interaction, which induce a coupling between axial motion and in--plane hot vibrational modes, and lead to a residual qubit--phonon coupling.
The description of such decoherence poses an involved theoretical problem, because of the large number of vibrational modes that participate in the process. 
However, it gets simplified due to the fact that the environment is in a gaussian state, which allows us to simplify the calculation of correlation functions that appear in finite--temperature, time--dependent perturbation theory. 
We show that in a range of parameters where axial confinement is large enough, the error induced in the quantum gate is very small. 
Furthermore, the adjustment of the gate time allows us to correct for the influence of the phonon environment, and decrease the gate error by more than one order of magnitude.

Let us consider a system of $N$ ions forming a 2D Coulomb crystal in a Penning trap. 
We study, for concreteness, the performance of a ``pushing gate'' \cite{Cirac.00} in the axial direction, an approach that has the advantage that single vibrational modes do not have to be resolved, and the gate can operate at finite temperature.  Since we are interested in estimating the consequences of decoherence, we can neglect finite size effects, and describe the crystal by a regular triangular lattice with periodic boundary conditions. Let us assign the ${\bf z}$ direction to the axis of the trap, such that ions occupy equilibrium positions in the ${\bf x}$--${\bf y}$ plane: 
\be 
\RR^0_\rr = (r_1 {\bf a}_1 \! + \! r_2 {\bf a}_2) d_0, \ \ \  r_j = 1, \dots , L,
\label{equilibrium.positions}
\ee
where ${\bf a}_1 = (1,0,0)$, ${\bf a}_2 = (1/2,\sqrt{3}/2,0)$, 
and $d_0$ is the distance between ions.  
The potential is given by a trapping term, plus the Coulomb repulsion:
\bea
V  &=&  V_{\textmd{Trap}} + V_{\textmd{Coul}} ,
\nonumber \\
V_{\textmd{Coul}} &=&
\frac{1}{2} \sum_{\rr,\ss} 
\frac{e^2}{|\RR^0_\rr \! - \! \RR^0_\ss + {\bf R}_\rr \! - \! {\bf R}_\ss|} .
\label{potential}
\eea 
$\RR_\rr = (X_\rr,Y_\rr,Z_\rr)$, are the ions' coordinates with
respect to the equilibrium positions, and $V_\textmd{Trap}$ is the
harmonic trapping potential, with frequencies $\omega_j$ in each
spatial direction, $j=$ $x,y,z$. Note that in a Penning trap,
(\ref{potential}) corresponds to the potential in a frame rotating
with the ion crystal \cite{note.magnetic.field}. 

In the harmonic approximation, $V_{\textmd{Coul}}$ is expanded up to second order in $\RR_\rr$, and the axial ($\bf z$) and in--plane (${\bf x}, {\bf y}$) modes are independent. The vibrational harmonic Hamiltonian is given by: 
\be
H^{(0)}_{\textmd{vib}} = \sum_{\lambda,\qq} \hbar \omega^\lambda_{\qq} a^\dagger_{\qq,\lambda} a_{\qq,\lambda}.
\label{harmonic.hamiltonian}
\ee
$\qq$  is the phonon wave--vector, and $\lambda$ runs over the three possible polarizations: $\lambda = z$ (axial modes), and the two in--plane modes ($\lambda = \ \parallel, \perp$) corresponding to longitudinal and transverse modes. Local displacements can be expressed in terms of collective coordinates: 
\bea
\RR_\rr = (1/\sqrt{N}) \sum_{\lambda,\qq} {\bf e}^\lambda_\qq  R^\lambda_\qq e^{i \qq \rr}, \nonumber \\
R^\lambda_\qq = \sqrt{\hbar/2 m \omega^\lambda_\qq} \left( a^\dagger_\qq + a_{-\qq} \right) .
\eea
where ${\bf e}^\lambda_\qq$ are the polarization vectors, which are eigenstates of the Fourier transform of the harmonic ion--ion interaction \cite{Dubin.93}: 
\bea
\Omega_\qq^{i,j} &=& 
\delta_{i,j} \ \omega_j^2 +  
\sum_\ss \frac{e^2}{m} (1 - \cos(\ss \qq) ) V_\ss^{i,j}, \nonumber \\
V_\ss^{i,j} &=& \frac{1}{|\RR^0_\ss|^3} \left( \frac{3 (\RR^0_\ss)_i (\RR^0_\ss)_j }{|\RR^0_\ss|^2} - \delta_{i,j} \right).
\label{potential.harmonic}
\eea
For each $\qq$, $\omega^\lambda_\qq$ are given by the eigenvectors of the $3 \times 3$ matrix $\sqrt{\Omega_{\qq}}$.

The ``pushing gate'' works by coupling the internal states of two ions, which we label $1$ and $2$, to the axial ($\bf z$) motion \cite{Cirac.00,pushing.gate}. 
An off--resonant standing--wave induces a force which displaces the position of ions $1$ and $2$, in a direction which depends on their internal state \cite{note1} (Fig. \ref{axial.scheme}):
\be
H_{\textmd{f}}(t) = \sum_{j = 1,2} F(t) Z_{\rr_j} \sigma^z_j ,
\label{axial.gate}
\ee
where $\sigma^z_j$ are operators acting on the internal states of ions at sites $\rr_j$.
To simplify our calculations, we consider the following
time--dependent force: $F(t) = F e^{- i \Gamma |t|/2}$ ($-\infty < t <
\infty$). However, our results are qualitatively valid for other pulse
shapes with gate time $1/\Gamma$. Trapping parameters are chosen such
that $\beta_z = e^2/m \omega^2_z d_0^3 \ll 1$ \cite{note1}. In this limit, ions
moving in the axial direction can be considered as independent
harmonic oscillators weakly coupled by the Coulomb interaction, akin
to the case of microtrap arrays. Furthermore, 
we will be in the adiabatic limit, defined here by:
\be
{\cal E}_z = 8 (\Gamma/\omega_z)^2 (F Z_0 / \hbar \omega_z)^2 (2
\bar{n}_z + 1) \ll 1 ,
\label{adiabaticity}
\ee
where $Z_0$ is the axial ground state size, and $\bar{n}_z$ is the mean axial phonon number. 
${\cal E}_z$ is, indeed, the error induced in the quantum gate by
nonadiabatic effects in the axial vibrational degrees of freedom \cite{pushing.gate}.
Under condition (\ref{adiabaticity}), internal states end up being decoupled from the axial motion after the gate, and follow the unitary evolution given by: 
\be
U_\textmd{g} = e^{- i \int d t J(t) \sigma^z_1 \sigma^z_2 }, 
\ J(t) = 2 \beta_z \left( \frac{F(t) Z_0}{\hbar \omega_z} \right)^2 \omega_z .
\label{gate.unitary}
\ee
In the following we consider the choice $\Gamma = J(0) \pi/8$, such that $U_\textmd{g}$ corresponds, up to local operations, to a sign gate. 
 
\begin{figure}
\resizebox{\linewidth}{!}{\includegraphics{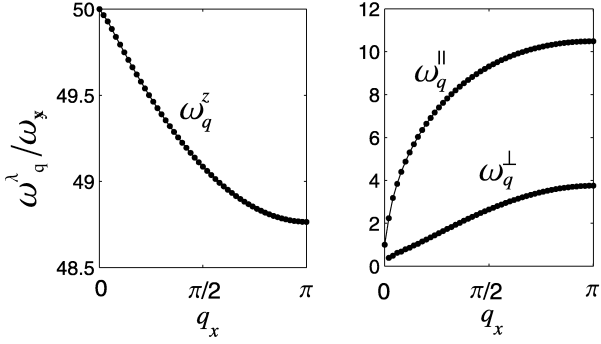}}
  \caption{Spectrum of the vibrational modes with wavevector $\qq$ along the $\bf x$ direction. $N=10^4$ ions, $\omega_z = 50 \omega_{\xy}$, and we consider periodic boundary conditions.}
\label{modes.spectra}
\end{figure}

Unfortunately, the axial motion is coupled to the in--plane modes by
anharmonic terms in $V_{\textmd{Coul}}$, being the lowest order ones
of the form $X Z^2$, $X^2 Z^2$, that is, quadratic in the axial
coordinates. 
Since resonances between axial and in--plane vibrational frequencies
lead to divergences in the correlation functions of these terms, the
effect of anharmonicities is reduced if
$\omega^{z}_{\qq} \gg 2 \ \omega^{\parallel,\perp}_{\qq}$. The axial
vibrational bandwidth is proportional to $\beta_z$, and, thus, this
condition can be imposed by choosing a thight enough axial confinement
(see \cite{note0}). In the case of $N$ = $10^4$ ions, $\omega_z
\approx 50 \ \omega_{\xy}$ 
is enough to ensure this condition (see Fig. \ref{modes.spectra}), 
so that we will assume this ratio in all the examples presented along this work.

Let us now quantify the loss of fidelity induced by anharmonic couplings. 
The force (\ref{axial.gate}) displaces the ions in the axial direction by $\bar{Z}(t) = 2 F(t) Z_0^2 / \hbar \omega_z$. In the limit (\ref{adiabaticity}),
ions 1 and 2 follow adiabatically the displacement induced by the force, so that one can neglect the fluctuations in the coordinates $Z_{\rr_j}$, and replace them by their ground state average, $\bar{Z}(t) \sigma^z_j$, when computing anharmonic corrections.  The anharmonic energy dependence on the ions' position results in the following coupling between internal states and in--plane modes:
\bea
H^{\textmd{dec}}_{\textmd{xy}}(t) &=&
H^\ah_\xy(t) \frac{1}{4} \left( \sigma_1^z - \sigma_2^z \right)^2 , \nonumber \\
H^\ah_\xy(t) &=& \bar{Z}(t)^2
\left(  {\bf A}^\textmd{T} \ \RR_{1,2} \ +  \RR_{1,2}^\textmd{T} {\bf B} \ \RR_{1,2} \right).
\label{anharmonic.coupling}
\eea
$\RR_{1,2}$ ($\RR_{1,2}^0$) is the vector given by the in--plane components of $\RR_{\rr_1} - \RR_{\rr_2}$ 
($\RR^0_{\rr_1} - \RR^0_{\rr_2}$). $\bf A$ and $\bf B$ are  third and fourth order anharmonic Coulomb interaction terms, respectively: 
\bea
{\bf A} = 3 \frac{e^2}{d_0^5}  \RR^0_{1,2} , \hspace{0.25cm} 
{\bf B} = 
\frac{3 e^2}{ d_0^5} 
\left(  {\bf 1} - \frac{5}{2} 
\frac{\RR^0_{1,2} \ ({\RR^0_{1,2}})^\textmd{T}}{d_0^2} \right) .
\label{anharmonic.coefficients}
\eea
Note that $H^{\textmd{dec}}_\xy$ excites the in--plane phonons depending on the ions' internal states, and, thus, entangles the qubits with the environment.

Assume that the internal states are initially in a pure state: 
$| \Psi \rangle = \sum_\alpha c_\alpha |\alpha \rangle$ 
($|\alpha \rangle  = |00 \rangle, |01 \rangle, |10 \rangle, |11 \rangle$), 
such that the initial density matrix of the total system is given by:
$\rho_i = | \Psi \rangle \langle \Psi | \otimes \rho_{\textmd{xy}}^0$, with $\rho^0_{\textmd{xy}}$, a thermal phonon state. 
After the action of the quantum gate, the final density matrix is:
\be
\rho_f = 
\sum_{\alpha,\beta} c_\alpha c^*_\beta 
\left( U_\textmd{g} |\alpha \rangle \langle \beta| U_\textmd{g}^\dagger \right) 
\left( U_{\textmd{xy}}^\alpha \rho_{\textmd{xy}}^0 (U^\beta_{\textmd{xy}})^\dagger \right),
\label{evolution}
\ee 
where $U^\alpha_{\xy}$, is the evolution operator of the in--plane
phonons, $U_\xy$, projected in the internal state $\alpha$. In (\ref{evolution})
we have assumed that the axial modes can be described classically in 
$H^{\textmd{dec}}_\xy$ due to the adiabaticity of the quantum gate,
such that we can factorize the evolution operator.
We define the (worst--case) reduced fidelity, ${\cal F}$, of the
quantum gate by the overlap of the qubits' reduced density matrix
obtained from (\ref{evolution}), with the qubit quantum state after a perfect gate, minimized over all possible two--qubit initial states.
Note that $U^{00}_\xy = U^{11}_\xy = U^{[0]}_\xy$, and 
$U^{01}_\xy = U^{10}_\xy = U^{[\ah]}_\xy$, where $U^{[0]}_\textmd{xy}$ and $U^{[\ah]}_\textmd{xy}$, are the in--plane phonon evolution operator in the absence and presence of anharmonic couplings, respectively. 
Thus, the fidelity is completely determined by the following complex quantity $\bar{\cal F}$: 
\bea
{\cal \bar{F}} 
&=&  
\tr {\big \{} | 01 \rangle \langle 00 | \rho_i  {\big \}} = 
\tr_\xy {\big \{} U_{\textmd{xy}}^{[\ah]} \rho_{\textmd{xy}}^0 {U_{\textmd{xy}}^{[0]}}^\dagger {\big\}} \nonumber \\
&=&  \tr_\xy {\bigg \{} {\cal T}\exp \left( -i \! 
\int_{-\infty}^\infty \hspace{-0.2cm} H^{\textmd{ah}}_{\textmd{xy}}(\tau) d \tau \right) \ \rho_{\textmd{xy}}^0  {\bigg \}},
\label{error}
\eea
which corresponds to the mean value of the evolution operator
$U_{\xy}^{[\ah]}$, in the interaction picture with respect to
$H^\ah_\xy$. 
Note that (\ref{error}) is similar to the expression found in the
theory developed in \cite{Prozen}.
The worst--case error is ${\cal E} = 1 - {\cal F} = (1 -\Re( \bar{\cal F} ))/2$. However, note that the fidelity can be improved, since, according to Eq. (\ref{anharmonic.coupling}), the spin--phonon coupling depends on the operator $(1 - \sigma^z_1 \sigma^z_2)$. For this reason, a correction of the gate time allows us to cancel the phase of $\bar{\cal{F}}$, and  
 define ${\cal E'} = (1 - |\bar{\cal F}|)/2$, which quantifies the worst--case error after the correct calibration of the gate duration.   

Eq. (\ref{error}) is a good starting point for perturbation theory, which can be carried out by expanding the time--ordered exponential. First of all, let us study the scaling of the terms appearing in $H^{\ah}_\xy$, and check whether a peturbative approach is indeed justified. The anharmonic coupling (\ref{anharmonic.coupling}) can be rewritten in terms of collective variables in the interaction picture:
\be 
H^{\ah}_{\xy}(\tau) = \sum_{\lambda, \qq} F^\lambda_\qq(\tau) R^\lambda_\qq(\tau) + 
\sum_{\lambda,\lambda',\qq,\kk} G^{\lambda \lambda'}_{\qq \kk}(\tau) R^\lambda_\qq(\tau) R^{\lambda'}_{\kk}(\tau),
\ee
According to Eqs. (\ref{gate.unitary}, \ref{anharmonic.coupling}, \ref{anharmonic.coefficients}), each term scales like:
\bea
\overline{F^\lambda_\qq R^\lambda_\qq} 
&\approx& J \frac{X_0}{d_0} \nonumber \approx \Gamma \left( \frac{X_0}{d_0} \right), \\
\overline{G^{\lambda,\lambda'}_{\qq,\kk} R^\lambda_\qq R^{\lambda'}_{\kk}} &\approx& J \left( \frac{X_0}{d_0} \right)^2 
\approx \Gamma \left( \frac{X_0}{d_0} \right)^2, 
\label{scaling}
\eea
where $X_0$ is the size of the ground state in the radial trapping
potential, and by $\overline{O}$, we mean the square root of the variance of the operator in
the ground state.
Since the evolution of the in--plane modes is governed by the in--plane trapping frequency, $\omega_{\xy}$, we expect that terms in perturbation theory scale according to $\Gamma/\omega_{\xy}$ in Eq. (\ref{scaling}). 
We study two sets of experimental parameters with $\omega_{\xy} = 20$
($200$) kHz, and $\omega_z = 1$ ($10$) MHz, which implies $d_0 = 46.8$ ($10.1$) $\mu$m. Let us consider $N = 10^4$ ions, such that both values lead to $\beta_z = 3.8 \cdot 10^{-3}$. Typical temperatures of $T =$ 1 mK can be reached after Doppler cooling, such that mean phonon numbers are $\bar{n}_z \approx $ $20$ ($2$) in the axial modes, and $\bar{n}_{\xy} \approx$ $10^3$ ($10^2$), in the center of mass in--plane modes. If we choose $(F Z_0 / \hbar \omega_z) = 0.234$, such that ${\cal E}_z < 10^{-5}$, we get $\Gamma/\omega_\xy =$ $5 \ 10^{-2}$. 
Taking into account that $X_0/d_0$ is small (as it should be, if the assumption of harmonic vibrational modes is valid), all terms are small in (\ref{scaling}). In particular, $X_0/d_0$ = $3.6 \cdot 10^{-3}$, and $5.0 \cdot 10^{-3}$, in the cases $\omega_\xy = 20$, and $200$ kHz, respectively.

To quantify precisely the error in the quantum gate, we proceed as
follows (see the Appendix for the details). 
We expand the time ordered exponential (\ref{error}) up to fourth
order in $H^\ah_\xy$, and keep all the terms up to order 
$\left( X_0 / d_0 \right)^4$.
Each contribution can be expressed in terms of time integrations of
correlation functions of collective coordinates $R^\lambda_\qq$. 
Resonances appear, for example, in those terms of second order in 
$G^{\lambda,\lambda'}_{\qq,\kk}$ which are of the form 
$\langle R^\lambda_\qq(t_1) R^{\lambda'}_\kk(t_1) 
R^\lambda_{-\qq}(t_2) R^{\lambda'}_{-\kk}(t_2) \rangle$
The problem gets numerically tractable by the application of Wick's theorem, which allows us to express real time high order correlation functions in terms of the two--operator correlation function. Note that Wick's theorem can be applied here, even when we are at finite temperatures, because of the statistical properties of the phonon thermal (gaussian) state. Finally, a summation of sets of diagrams up to infinite order is possible by means of the linked cluster theorem, such that (\ref{error}) takes finally the form of the exponential of low order irreducible diagrams \cite{Negele}.  

\begin{figure}
  \resizebox{\linewidth}{!}{%
    \includegraphics{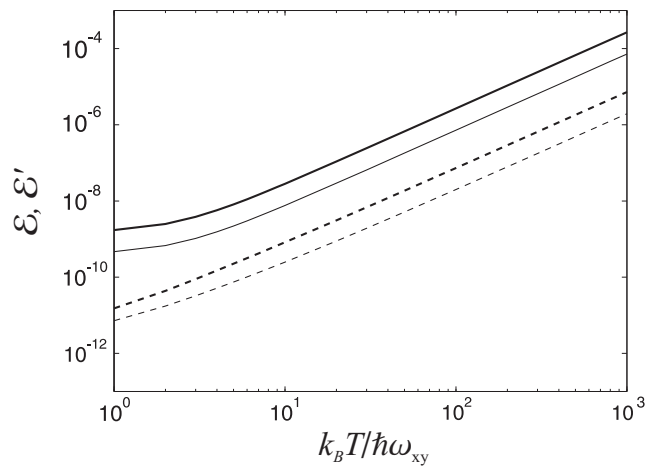}}
  \caption{Continuous lines: ${\cal E}$ (error without phase correction). Dashed lines: ${\cal E}'$ (error with correction of the gate time. Thick lines and thin lines, correspond to $\omega_\xy = 20$ kHz and $\omega_\xy = 200$ kHz, respectively. $N$ = $10^4$ ions and $\Gamma/\omega_\xy$ = 0.05.}
\label{errT}
\end{figure}

The results of our calculation are presented in Fig. \ref{errT}, where we show our results for the two sets of experimental parameters discussed above.
Note that at high temperatures, like those ones that occur in current
experiments with Penning traps, we get a quadratic dependence with
temperature, which can be explained by the dominant contribution of
terms that are second order in the coupling
$G^{\lambda,\lambda'}_{\qq,\kk}$.  
The correction of the phase by adjustment of the gate time allows us
to reduce the error by more than one order of magnitude. 
Note that even with the highest temperatures considered here, which
correspond to the limit of Doppler cooling, anharmonic terms induce
very small errors. With the range of parameters considered in this
work, we get rates $\Gamma =$ $1$, $10$ kHz. A limitation in the
number of quantum gates due to heating, reduces the number of gates to
a maximum of $\approx$ $10^2$, with present heating rates, 
however, this quantity is amenable to be
improved by increasing the quality of the vacuum in the trap \cite{Bollinger.heating} . 

Finally, our proposal could also be used for the quantum simulation of interacting spin--systems, by applying a state dependent force to all the ions at the same time. In this way, as shown in Ref. \cite{spin.simulator}, an antiferromagnetic Ising interaction is induced between the internal states, which behave like effective spins. 
If we add a transverse field of the form $(\Omega/2)\sum_\rr \sigma^x_\rr$, by means of a global carrier transition, then this experimental set--up allows us to study the rich phenomenology of quantum frustration in triangular lattices \cite{ising.frustration}. Note that our analysis of the decoherence induced by low--energy vibrational modes, also implies the viability of this approach, since the effective spin--spin interactions are an always--on version of the qubit--qubit coupling induced during the quantum gate.


We thank J. Bollinger, D. Leibfried and  M. Aguado for interesting discussions. Work supported by CONQUEST, SCALA, and Marie Curie under contract MEIF--2004--010350.

\appendix

\section{Properties of a Wigner 2D Crystal}
{\it (In this appendix, we describe in more detail the properties of a
  2D Wigner
crystal, as well as the calculation of the decoherence in the quantum
gate due to coupling with low--energy vibrational modes. For the sake
of readability, we repeat a few of the equations already presented in
the main text.)}

%
%

We express the potential in the rotating frame in terms of the equilibrium positions $\RR^0_\rr$,
and the coordinates of the ions with respect to the equilibrium positons,
$\RR_\rr$:
\begin{eqnarray}
V &=& V_{\textmd{Coul}} + V_{\textmd{Trap}}, \nonumber \\
\nonumber \\
V_{\textmd{Coul}} &=&
\frac{1}{2} \sum_{\rr,\ss} 
\frac{e^2}{|\RR^0_\rr \! - \! \RR^0_\ss + {\bf R}_\rr \! - \! {\bf
    R}_\ss|}, \nonumber \\
V_{\textmd{Trap}} &=&
\frac{1}{2} m \omega^2_z \sum_\rr \!  \left(Z^0_\rr + Z_\rr \right)^2 + \nonumber \\
&+&
\frac{1}{2} m \omega^2_r \sum_\rr \! 
\left( \left( X_\rr^0 \! + \! X_\rr \right)^2 \! + \! 
       \left( Y_\rr^0 \! + \! Y_\rr \right)^2 \right). 
\label{potential.again}
\end{eqnarray}
In the case $\omega_\zz \gg \omega_\xy$, the Coulomb crystal is a single plane,
which corresponds to a 2D Wigner crystal. Under these conditions, ions
arrange themselves in the triangular lattice generated by
(\ref{equilibrium.positions}). 

The theoretical description of this system becomes much simpler if
we consider periodic boundary conditions, something that is a good
approximation, since we deal with large number of particles.
The equilibrium positions in a 2D Wigner crystal are given by a
triangular lattice generated by:
\begin{equation}
{\bf a}_1 = (1, 0, 0), 
\hspace{0.5cm}
{\bf a}_2 = (1/2 ,\sqrt{3}/2, 0) .
\end{equation} %
The reciprocal lattice is generated by the following vectors: 
\begin{equation}
{\bf b}_1 = (1, - 1/\sqrt{3}, 0) , 
\hspace{0.5cm}
{\bf b}_2 = (0 , 2/\sqrt{3}, 0), 
\end{equation} 
The basis of the reciprocal lattice is determined by
the relations: 
$({\bf a})_i \cdot ({\bf b})_j = \delta_{i,j}$. 
They are the allowed directions of propagation given by the periodicity of the
ion lattice. 

Since we are considering a lattice with regular spacing, we can easily
determine the ions' equilibrium positions. In this simple case, this
task is reduced to finding the equilibrium distance between ions,
$d_0$, which we prefer to express in terms of the adimensional
parameter $\beta_\xy$:
\begin{eqnarray} 
\beta_\xy(L) &=& \frac{2 e^2}{m \omega_\xy^2 d_0^3} = \frac{4
  C_1(L)}{L^2 C_2(L)}, 
\nonumber \\
C_1(L) &=& \sum_\ss |\RR^{0}_\ss|^2 , 
\nonumber \\ C_2(L) &=& \sum_\ss 1/|\RR^0_\ss| .
\end{eqnarray} 

To determine dynamical properties of the Coulomb crystal, we study now
the normal vibrational modes, which diagonalize the motion Hamiltonian in the
harmonic approximation. Remember that these ones are given by
plane--waves of the following form:
\begin{eqnarray} 
\RR_{\qq} &=& \frac{1}{L} \ \sum_\rr e^{i \qq \rr} \RR_{\rr} , 
\nonumber \\ 
\PP_{\pp} &=& \frac{1}{L} \ \sum_\rr e^{- i \pp
\rr} \PP_{\rr} .
\end{eqnarray} 
Let us keep in mind that these are not hermitian
operators $(\RR_\qq)^\dagger = \RR_{-\qq}$. 
They satisfy the commutation relations: 
$[\RR^i_{\qq}, \PP^j_{\pp}] = i \delta_{\qq,\pp} \delta_{i,j}$, such that the harmonic
Hamiltonian is finally written like this: 
\begin{equation} 
H^{(0)}_{\textmd vib} = \sum_{\qq,i,j}
\frac{1}{2} m \Omega^{i,j}_\qq (\RR_{\qq})^i (\RR_{-\qq})^j 
+ \sum_{\qq} \frac{1}{2 m} \PP_{\qq} \PP_{-\qq} .  
\end{equation} 
Where the dispersion
relation is the Fourier transform of the relative potential, see
Eq. (\ref{potential.harmonic}). Allowed phonon wavevectors are $\qq = q_1 {\bf b}_1 +
q_2 {\bf b}_2$, that is, it is a vector of the reciprocal lattice (otherwise
the plane--waves do not satisfy the orthogonality relations on the
triangular lattice). 
Local coordinates can be finally written like:
\begin{eqnarray} 
\RR_\rr &=& \frac{1}{L} \sum_{\lambda,\qq} e^{-i \qq \rr}
 {\bf e}^\lambda_\qq \sqrt{\frac{\hbar}{2 m
\omega^\lambda_\qq}} \left( a^\dagger_{\lambda,\qq} + a_{\lambda,-\qq}
\right) ,
\nonumber \\ 
{\bf P}_\rr &=& \frac{i}{L} \sum_{\lambda,\qq}
e^{-i \qq \rr}  {\bf e}^\lambda_\qq  \sqrt{\frac{m \hbar
\omega^\lambda_\qq}{2}} \left( a^\dagger_{\lambda,\qq} -
a_{\lambda,-\qq} \right) ,
\end{eqnarray}
where ${\bf e}^\lambda_\qq$ determine the direction of vibration of
the mode, and the index $\lambda$ can take three values which
correspond to longitudinal or transverse modes in the in--plane
direction, and vibration perpendicular to the Coulomb crystal plane.

The anharmonic corrections to the vibrational Hamiltonian that are
relevant for us, are those which couple the axial ({\bf z}) to
the in--plane motion ({\bf x}, {\bf y}), because they will lead
eventually to the qubit--phonon couplings presented in Eq. (\ref{anharmonic.coupling}).  
Let us writte them explicitly:
\begin{eqnarray}
H_{\textmd{ah}} &=&
\frac{1}{2} 
\sum_{\stackrel{\rr,\ss}{i=x,y}} \! \!
A^i_{\rr,\ss} \ (\RR_{\rr,\ss})^i
(\RR_{\rr,\ss})^z+
\nonumber \\
&+&
\frac{1}{2} \sum_{\stackrel{\rr,\ss}{i,j=x,y}} \! \! 
B^{i,j}_{\rr,\ss} \ (\RR_{\rr,\ss})^i (\RR_{\rr,\ss})^j
(\RR_{\rr,\ss})^z .
\label{anharmonic}
\end{eqnarray}
where $\RR_{\rr,\ss} = \RR_\rr - \RR_\ss$. 
Note that we include in $H_\ah$ only those terms that have the form 
$H_{\ah} \approx X \ Z^2 + X^2 \ Z^2$. The anharmonic coupling constants given by:
\begin{eqnarray}
A^i_{\rr,\ss} &=& 
\frac{3}{4} \frac{e^2}{d_0^5}
\frac{\left(\RR^0_{\rr,\ss}\right)^i}{|\RR^0_{\rr,\ss}|^4}, 
\nonumber \\
B^{i,j}_{\rr,\ss} &=& 
\frac{3}{8} \frac{e^2}{d_0^5}
\left(
2 \delta_{i,j} - 
5 \frac{\left(\RR^0_{\rr,\ss}\right)^i \left(\RR^0_{\rr,\ss}\right)^j}{|\RR^0_{\rr,\ss}|^5}, 
\right).
\label{anharmonic.coupling.constants}
\end{eqnarray}
Again, we define relative coordinates $\RR^0_{\rr,\ss} = \RR^0_\rr - \RR^0_\ss$.

Since we want to induce a quantum gate between ions by means of an
internal--state dependent force in the axial (${\bf z}$) direction,
the position of the ions will depend, in the adiabatic limit of the
gate, on the internal state. This induces a coupling between qubits
and in--plane modes, through the dependence of (\ref{anharmonic}) on
$\RR^z$. To get an explicit form for this coupling, let us first
introduce the formalism for describing a pushing gate between to
nearest--neigbors in the Coulomb crystal.

\section{Pushing gate in the axial direction}

Our goal now is to study the performance of the pushing gate under
the presence of anharmonic terms. 
First of all, let us collect all the terms that describe our system:
\begin{eqnarray}
H(t) &=& H_0(t) + H_\ah, \nonumber \\ 
H_0(t) &=& H^{(0)}_{\textmd{vib}} + H_{\textmd{f}}(t) 
\end{eqnarray}
$H_0(t)$ is the harmonic part of the Hamiltonian, and can be solved
exactly. We include in it the vibrational Hamiltonian:
\begin{equation}
H^{(0)}_{\textmd{vib}} 
= \sum_{\lambda,\qq} \hbar \omega_{\lambda,\qq} a^\dagger_{\lambda,\qq} a_{\lambda,\qq},
\label{hamiltonian.vibrational}
\end{equation}
as well as the force in the axial direction:
\begin{equation}
H_\textmd{f} = \sum_{j=1,2} F(t) Z_{\rr_j} \sigma^z_j.
\label{hamiltonian.whole.system}
\end{equation}
Under the set of conditions considered in this work, the error in the
quantum gate is the sum of two independent contributions: {\it (i)}
nonadiabatic effects due to the excitation of phonons in the axial
direction (common to any implementation of the pushing gate), and {\it (ii)} the
errors induced by anharmonic couplings. Our strategy will be, first,
to solve $H_0(t)$, then to study $H_{\ah}$ in the interaction picture
with respect to $H_0(t)$, and finally to evaluate the effect of the
anharmonic couplings by doing perturbation theory.

For later convenience, we choose the following temporal profile for
the amplitude of the off--resonant standing wave:
\begin{equation}
F(t)^2 = F^2 P(t), \hspace{1.cm} P(t) = e^{- \Gamma |t|} .
\label{noch.einmal.force}
\end{equation}
We define an interaction picture with respect
to $H_{\textmd{ah}}$, which has the only unusual feature that $H_0(t)$ is
time--dependent: 
\begin{eqnarray} 
U_{\textmd I} (t) &=& U_0^{\dagger}(t) U(t) , 
\nonumber \\
\partial_t U_{\textmd I} (t) &=& - \frac{i}{\hbar} 
U^\dagger_0(t) H_{\ah} U_0(t),
\end{eqnarray} 
where $U(t)$, $U_0(t)$, are the evolution operators corresponding to
$H(t)$, $H_0(t)$, respectively, and $U_{\textmd{I}}(t)$ is the evolution
operator in the interaction picture. 

We consider the limit $\beta_z \ll 1 $, that is, the Coulomb
interaction between ions is small with respect to the trapping
potentials. 
The problem could be easily extended to the general case,
but remember that this condition has to be fulfilled to avoid
crossings between in--plane modes and the axial motion of the ions.
In this limit, the evolution operator consists of qubit--qubit couplings, as well as
the displacement induced by the force (which can be considered
independently for each ion):
\begin{eqnarray}
U_0(t) &=& U_\textmd{g}(t)
e^{-i H^{(0)}_{\textmd{vib}} t}
e^{\sum_{j=1,2} \left( \eta(t)^* a_{\zz,j} -  \eta(t) a^\dagger_{\zz,j}
\right) \sigma^\zz_j
},
\nonumber \\
U_\textmd{g}(t) &=& 
e^{-i \int_{-\infty}^t J(\tau) \sigma^\zz_1 \sigma^\zz_2 d \tau } ,
\label{unitary.0}
\end{eqnarray}
where $\eta(t)$ determines the time--dependent displacement of each
ion's position:
\begin{equation}
\eta^*(t) = (-i) \int_{-\infty}^t \frac{F(\tau) Z_0 }{\hbar}
e^{- i \omega_\zz \tau} d \tau,
\label{eta}
\end{equation}
Furthermore, we consider the adiabatic limit, $\Gamma \ll \omega_z$, 
in which the coupling $J(t)$ is given by:
\begin{equation}
J(t) = 2 \beta_z \left( \frac{F(t) Z_0}{\hbar \omega_z} \right)^2
\hbar \omega_\textmd{z},
\label{interaction.strength}
\end{equation}
and the ions $1$ and $2$ are displaced by the state-dependent force in the
following way:
\begin{eqnarray} 
U_0^\dagger(t) Z_j U_0(t) &=& 
Z_j(t) + 2 \eta(t) Z_0 \sigma^z_j 
\end{eqnarray} 
with $\eta(t)$ given by:
\begin{equation} 
\eta(t) = \frac{F(t) Z_0}{\hbar \omega_z} 
\label{eta.adiabatic}
\end{equation} 
The coupling of the qubits with the in--plane vibrational modes is
obtained formally by studying the evolution of $H_{\ah}$ in the
interaction picture:
\begin{eqnarray}
H^\textmd{I}_\ah(t) &=&
U_0^\dagger(t) H_\ah U_0(t) \approx \nonumber \\
& & \hspace{-2.0cm}  - 4 \eta(t)^2 Z_0^2 \ 
\left(
(\sigma^\zz_1 - \sigma^\zz_2)^2 
A^i_{\rr_1,\rr_2} \sum_i (\RR_{\rr_1,\rr_2})^i + \right.
\nonumber \\
& & \hspace{-1.25cm}  \left. \sum_{i,j}
B^{i,j}_{\rr_1, \rr_2} \
(\RR_{\rr_1,\rr_2})^i 
(\RR_{\rr_1,\rr_2})^j \right) = H^\textmd{dec}_\xy(t).
\label{anharmonic.interaction.picture}
\end{eqnarray}
In the last equation we have neglected all the terms that depend on
the operators $Z_j$, because they give negligible contributions in the
adiabatic limit of the quantum gate. In this limit,
Eq. (\ref{anharmonic.interaction.picture}) reduces to a coupling
between in--plane vibrational modes and qubits, which does not involve any quantum
dynamics in the axial motion. Recall that
Eq. (\ref{anharmonic.interaction.picture}) defines
$H_\xy^\textmd{dec}$, that is, the coupling that we have used in the
main text.

We have already stated the problem in terms that are suitable to study
the fidelity of the pushing gate. For this task, let us assume that the
internal states are initially in a pure state:
\begin{equation} 
| \Psi \rangle =
\sum_\alpha c_\alpha | \alpha \rangle, \ \ 
| \alpha \rangle = | 00 \rangle, | 01 \rangle, | 10 \rangle, 
| 11 \rangle . 
\end{equation} 
In this appendix we are describing both the error induced by
in--plane modes, and the error due to nonadiabatic corrections in the
axial motion, so that we  study the density matrix of the whole
system (axial modes, in--plane modes, and qubits). Let us consider, an
initial density matrix for the system, which describes the initial
qubit pure state, and thermal phonon states $\rho^0_\textmd{z}$, $\rho^0_\xy$:
\begin{equation}
\rho_i =  | \Psi \rangle \langle \Psi | \otimes \rho^0_{\textmd{z}} 
\otimes \rho_{\textmd{xy}}^0
\label{initial.density.matrix}
\end{equation}
After the action of the quantum gate, the final density matrix is:
\begin{equation}
\rho_f = 
\sum_{\alpha,\beta} c_\alpha c^*_\beta 
\left( U_\textmd{g} |\alpha \rangle \langle \beta| U_\textmd{g}^\dagger \right) 
\left( U_{\textmd{z}}^\alpha \rho_{\textmd{z}}^0 (U^\beta_{\textmd{z}})^\dagger \right)
\left( U_{\textmd{xy}}^\alpha \rho_{\textmd{xy}}^0
  (U^\beta_{\textmd{xy}})^\dagger \right) .
\label{final.density.matrix}
\end{equation}
In Eq. (\ref{final.density.matrix}), $U_\textmd{z}^\alpha$,
$U_\textmd{xy}^\alpha$, are the unitary evolution of the vibrational
modes in directions ${\bf z}$, and ${\bf x}$--${\bf y}$,
respectively, proyected into the qubits' quantum state $\alpha$, that
is:
\begin{eqnarray}
U^\alpha_\textmd{z}(t) &=& 
e^{\sum_{j = 1,2} 
\left( \eta(t)^* a_{\textmd{z},j} - \eta(t) a^\dagger_{z,j}  \right)
\langle \alpha | \sigma_j^z | \alpha \rangle }
e^{-i H^0_\textmd{z} t} ,
\nonumber \\
U^\alpha_\textmd{xy}(t) &=& 
e^{-i H_\xy^0 t} {\cal T} \exp \left(
\int_{-\infty}^t \hspace{-0.25cm}
\langle \alpha | H^\textmd{I}_\ah(\tau)  |
 \alpha \rangle   d \tau  \right).
\label{definitions.unitary.zxy}
\end{eqnarray}
where we have used the time--ordered exponential ${\cal T} \exp$, to
express the evolution operator in the interaction picture with respect
to $H_\ah$. We define ${\cal F}$, the reduced fidelity, as the overlap
of the qubits' reduced fidelity obtained from $\rho_f$, with a pure qubit--state after the action of the unitary operation, $U_\textmd{g}$: 
\begin{equation}
{\cal F} = \sum_{\alpha,\beta} |c_\alpha|^2 |c_\beta|^2
{\cal F}^{\alpha,\beta}_\textmd{z} {\cal F}^{\alpha,\beta}_\xy
\end{equation}
where we have defined the following quantities:
\begin{eqnarray}
{\cal F}^{\alpha,\beta}_\textmd{z} &=& 
\tr_\textmd{z}
\left( 
U_{\textmd{z}}^\alpha \rho_{\textmd{z}}^0
(U^\beta_{\textmd{z}})^\dagger \right) ,
\nonumber \\ 
{\cal F}^{\alpha,\beta}_\xy &=&
\tr_\xy
\left( U_{\textmd{xy}}^\alpha \rho_{\textmd{xy}}^0
  (U^\beta_{\textmd{xy}})^\dagger \right) .
\label{fidelity}
\end{eqnarray}
${\cal F}_\textmd{z}$ includes the decoherence induced in the quantum
gate by the finite gate rate, $\Gamma$, something that is general to any quantum gate that relies on the adiabatic displacement of vibrational modes (see, for example, \cite{pushing.gate}). Its value can be easily evaluated:
\begin{equation}
{\cal F}^{\alpha,\beta}_\textmd{z} = 
e^{- |\eta_{\textmd{ND}}|^2 (\bar{n}_\textmd{z} + 1/2) \left( \left( \langle \sigma^z_1
\rangle_\alpha - \langle \sigma^z_1 \rangle_\beta \right)^2 + 
\left( \langle
\sigma^z_2 \rangle_\alpha - \langle \sigma^z_2 \rangle_\beta \right)^2 \right)}
.  
\label{fidelity.z}
\end{equation}
where $\eta_{\textmd{ND}}$ includes the corrections due to nonadiabaticity, namely:
\begin{equation}
\eta_{ND} = -2 i \frac{F Z_0}{\hbar \omega_z} \frac{\Gamma}{\omega_z}.
\label{nonadiabatic}
\end{equation}
where $Z_0$ is the ground state size in the axial trapping potential.
The contribution to the error can be estimated from
(\ref{fidelity.z}). For this, we consider that we are in the limit of
small errors, and minimize ${\cal F}_\zz$ with respect to all the
possible initial states, to get the worst--case error, which is given by:
\begin{eqnarray} 
{\cal E}^z &\approx& 4 \eta^2 (\bar{n} + 1/2) \approx 16 \ \left(
\frac{\Gamma}{\omega_z} \right)^2 \left( \frac{F Z_0}{\omega_z}
\right)^2 .
\label{error.z}
\end{eqnarray} 
The main point of this work is the calculation of the decoherence that
is inherent to this system, that is, the one induced by the coupling
of the qubits to the in--plane motion. Since this one poses a more
complicated problem, we will be dedicated to it during the next sections to it
\section{Decoherence induced by coupling to in--plane vibrational modes}

Let us see how to express ${\cal F}_\xy$ in a form that is suitable
for doing perturbation theory in the anharmonic couplings. First of
all, note that the only matrix elements  
$\langle \alpha | H^I_\ah | \alpha \rangle$  which are different
from zero are those with 
$| \alpha \rangle  = | 0 1 \rangle, | 1 0 \rangle$:
\begin{equation}
H_\xy^\ah = \langle \alpha | H_\xy^\textmd{dec} | \alpha \rangle
\label{anharmonic.decoherence}
\end{equation}
The only quantity that we have to evaluate to
calculate the contribution to ${\cal F}$ in (\ref{fidelity}) is the
following one: 
\begin{equation}
\bar{{\cal F}}_\xy = \tr_\xy 
\left( {\cal T} \exp \left( \int_{-\infty}^t H_\xy^\ah(\tau) d \tau  \right) \right)
\label{fidelity.xy}
\end{equation}
Note that $\bar{{\cal F}}_\xy$ is named simply as $\bar{\cal F}$ in
the main text, because we focus there on the decoherence induced by
coupling to $x$--$y$ vibrational modes only.

We  assume that the ion 
$1$ is at the position $\RR^0_{\rr_1} = (0,0)$, and 
$2$ is at $\RR^0{\rr_2} = (1,0)$, so that
the following substitution allows us to express the coupling terms as
a function of normal modes: 
\begin{equation} 
\left( \RR_{\rr_1} \right)^i - 
\left( \RR_{\rr_2} \right)^i 
= \frac{1}{L} \sum_{\qq, \lambda}
\left( {\bf e}^\lambda_\qq \right)^i (1 - e^{i q_1}) R^{\lambda}_\qq 
\end{equation}
We rewrite here $H^\ah_\xy$ in a form that is more suitable to study the
scaling of the error with the different parameters of the problem: 
\begin{eqnarray}
H^\ah_\xy(\tau) &=& 
{\bar F} P(\tau)
\frac{1}{L} \sum_{\lambda, \qq}
F_\qq^\lambda \tilde{R}^\lambda_\qq (\tau) + 
\nonumber \\
& & \frac{1}{L^2}
{\bar G} P(\tau) \! \! \! \! 
\sum_{\qq,\kk,\lambda,\lambda'} \! \! \! \!
G^{\lambda,\lambda'}_{\qq,\kk} 
\tilde{R}^\lambda_\qq (\tau) \tilde{R}^{\lambda'}_\kk (\tau) ,
\nonumber \\
F^\lambda_\qq &=& 
{\bf e}^x_{q,\lambda} (1 - e^{i q_x}) , 
\nonumber \\
G^{\lambda,\lambda'}_{\qq,\kk} &=& 
\sum_{i,j}
\left(2 \delta_{i,j} - 5 \delta_{i,x}
\delta_{j,x} \right) 
\nonumber \\
&& \hspace{1.cm} {\bf e}^i_{\qq,\lambda} {\bf e}^j_{\kk,\lambda'}
(1-e^{i q_x})(1-e^{i k_x}) . 
\label{hamiltonian.ah.xy}
\end{eqnarray}
We define: $\eta(t)^2 = \eta^2 P(t)$, where $P(t)$ is defined in
Eq. (\ref{noch.einmal.force}), and $\eta = \eta(0) = F Z_0 / \hbar
\omega_z$. 
In (\ref{hamiltonian.ah.xy}), we have defined the adimensional
coordinates:
\begin{equation}
\tilde{R}^\lambda_\kk = R^\lambda_\kk/X_0
\end{equation} 
where we have defined the ground state size in the radial trapping
potential, $X_0 = \sqrt{\hbar/(2 m \omega_r)}$. The spin--phonon coupling constants are: 
\begin{eqnarray}
\bar{F} &=& - 16 \eta^2
\bar{z}^2 X_0 \frac{3 e^2}{4 d_0^4} = - 6 \beta_z \eta^2 \omega_z
\left( \frac{X_0}{d_0} \right) , 
\nonumber \\ 
\bar{G} &=& - 16
\eta^2 \bar{z}^2 (X_0)^2 \frac{3}{8} \frac{e^2}{d_0^5} = - 3 \beta_z
\eta^2 \omega_z \left( \frac{X_0}{d_0} \right)^2.
\label{F.G} 
\end{eqnarray} 
The advantage of
expression (\ref{F.G}) is that it shows explicitly the scaling of
the coupling of the qubits with the in--plane vibrational modes in
terms of the relevant quantities of the problem.

In the Taylor expansion of (\ref{fidelity.xy}), only
time--ordered terms appear. 
Even when we are at finite--temperature, 
Wick's theorem applies \cite{Negele}, something that can be
shown by means of the path integral formalism, or by studying directly
the generating functional of the correlation functions (see the last section of
this appendix). In this way, one can express every term in perturbation theory as
integrations of the contractions of the theory, which we define here
in the following way:
\begin{eqnarray}
& & D^{\lambda,\lambda'}_{\qq,-\kk}(1,2) = 
D^{\lambda,\lambda'}_{\qq,-\kk}(t_1 - t_2), 
\nonumber \\ 
& & D^{\lambda,\lambda'}_{\qq,-\kk}(\tau) = 
\langle {\cal T} \{ \tilde{R}^\lambda_\qq (\tau)
\tilde{R}^{\lambda'}_{-\kk} (0) \}  \rangle
\nonumber \\
& &  
= \frac{\omega_\xy}{\omega^\lambda_\qq} \left( n_{\qq,\lambda} e^{i \omega_\qq \tau} + 
n_{\qq,\lambda} e^{- i \omega_\qq^\lambda \tau} + \right.
\nonumber \\
& & 
\hspace{0.75cm} 
\left. 
\theta(\tau)  e^{- i \omega^\lambda_\qq \tau} + 
\theta(-\tau) e^{i \omega_{\lambda,\qq} \tau } \right) 
\delta_{\qq,-\kk} 
\delta_{\lambda,\lambda'} .
\end{eqnarray} 
where we made explicit that in the limit of high
temperatures time--order does not matter and our task should be
infinitely simplified.  

When calculating terms in perturbation theory,
\begin{equation} 
D^{\lambda, \lambda}_{\qq,-\qq} (\tau)
= \frac{1}{2 \pi} 
\int_{-\infty}^{\infty} 
e^{i \omega \tau} D^{\lambda,\lambda}_{\qq,-\qq}(\omega) d \omega .  
\label{free.propagator}
\end{equation} 
We will make use the following identity: 
\begin{equation} 
e^{i \omega^\lambda_\qq \tau} \theta(\pm \tau) = 
\frac{\pm 1}{2 \pi i} \int_{-\infty}^{\infty} 
\frac{e^{i \omega \tau}}{\omega - \omega^\lambda_\qq \mp i \delta} d \omega .  
\end{equation}
So that: 
\begin{eqnarray} 
D^{\lambda,\lambda}_{\qq,-\qq} (\omega) &=&
\frac{\omega_\xy}{\omega^\lambda_\qq} {\bigg (}
\nonumber \\
& &
\hspace{-1.0cm}
n_{\lambda,\qq} 
2 \pi \delta(\omega - \omega^\lambda_\qq)
+ n_{\lambda,\qq} 2 \pi \delta(\omega + \omega^\lambda_\qq)
\nonumber \\
& & \hspace{2.0cm}
+ \frac{2 i \omega^\lambda_\qq}
{\omega^2 - (\omega^\lambda_\qq)^2 + i \delta} {\bigg )} .  
\label{phonon.propagator}
\end{eqnarray}
The reader is going to find out the reason for our choice of
exponential pulses: their Fourier transform is a Lorentzian, which is
better suited for integrations with the phonon propagator in 
Eq. (\ref{phonon.propagator}),
\begin{eqnarray} 
P(t) &=&  \frac{1}{2 \pi} \int d\omega e^{i \omega t} P(\omega) , 
\nonumber \\ 
P(\omega) &=& \frac{2 \Gamma}{\omega^2 + \Gamma^2} .
\end{eqnarray}
The energy scales of this problem are $J$, $\Gamma$, $\omega_r$,
and $T$. $J$ and $\Gamma$ are related by the requirement that a sign
gate is realized by the coupling to the axial modes, 
that is $J = \frac{\pi}{8} \Gamma$, something that allows us to rewrite the
coupling constants in a self--consitent way: 
\begin{eqnarray} 
\bar{F} = -
\frac{3}{4} \pi \left( \frac{X_0}{d_0} \right) \Gamma , 
\nonumber\\ 
\bar{G} = - \frac{3}{8} \pi \left( \frac{X_0}{d_0} \right)^2 \Gamma .  
\end{eqnarray} 
Thus, it is clear from the previous expression, that $\bar{F}$,
$\bar{G}$ are small, both if $X_0/d_0 \ll 1$, and in the case in which
the quantum gate is also adiabatic with respect to the in--plane
motion. However, the scaling of the coupling constants with these
parameters does not allows us to extract definitive conclusions about
the decoherence, because the energies of the vibrational modes are
strongly corrected by the Coulomb energy (see Fig. \ref{modes.spectra}).

\section{Perturbation theory}
In this section, we present the calculation of the contributions to
the expansion of ${\cal F}_\xy$ which are of fourth order in
$\left( X_0 / d_0 \right)$, that is, up to 
${\cal O}({\bar F}^4)$, 
${\cal O}({\bar G}^2)$, 
${\cal O}({\bar G} {\bar F}^2)$. 

We have to evaluate time ordered averages of products of operators of
the form:
\begin{equation}
\langle {\cal T} \{
R^{\lambda_1}_{\kk_1}(t_1) \dots R^{\lambda_n}_{\kk_n}(t_n)
 \} \rangle .
\label{multi.time.correlation}
\end{equation}
The average $\langle \rangle$ is understood as a thermal average over
the in--plane vibrational modes. 
The calculation of time ordered correlation functions becomes simplified by the use of Wick's
theorem. Even when it is usually stated in terms of an
operator identity that is useful only at zero temperature, Wick's
theorem can be applied to finite temperature, by using the fact that
a thermal state is gaussian, and thus correlation functions  can
be obtained as functional derivatives of a gaussian functional. 
We give more details of this point in the following
subsection. For the moment, let us recall that Wick's theorem states
that (\ref{multi.time.correlation}) can be expressed as a sum of
products of two--time correlation functions, 
each of them corresponding to a possible pairing of the operators:
\begin{eqnarray}
& &
\langle {\cal T} \{
R^{\lambda_1}_{\kk_1}(t_1) \dots R^{\lambda_n}_{\kk_n}(t_n)
 \} \rangle = \nonumber \\ 
& &
D^{\lambda_1,\lambda_2}_{\kk_1,\kk_2}(t_1 \! \!  - \! \! t_2)
D^{\lambda_3,\lambda_4}_{\kk_3,\kk_4}(t_3 \! \!  - \! \! t_4) \dots
D^{\lambda_{n-1},\lambda_n}_{\kk_{n-1},\kk_n}(t_{n-1} \! \!  - \! \!
t_n) + \nonumber \\
& &
D^{\lambda_1,\lambda_3}_{\kk_1,\kk_3}(t_1 \! \!  - \! \! t_3)
D^{\lambda_2,\lambda_4}_{\kk_2,\kk_4}(t_2 \! \!  - \! \! t_4) \dots
D^{\lambda_{n-1},\lambda_n}_{\kk_{n-1},\kk_n}(t_{n-1} \! \!  - \! \!
t_n) + \nonumber \\
& & \dots \textmd{(all possible pairings)} .
\label{Wick.theorem}
\end{eqnarray}
Note that Wick's theorem allows us to
apply the cluster theorem to the expansion of the time order
exponential in Eq. (\ref{fidelity.xy}), that is:
\begin{equation}
\bar{\cal F}_\xy = \exp \left( {\cal E}^{(1)} + {\cal E}^{(2)} + \dots \right),
\label{cluster.theorem}
\end{equation}
where ${\cal E}^{(n)}$ are disconnected contributions only, that is,
terms in which the time integrations cannot be factorized into
separate independent time integrations.
\subsection{First order in $H^\ah_\xy$} 
The lowest order contribution is given by:
\begin{eqnarray} 
{\cal E}^{(1)} &=& (-i) 
\frac{\bar{G}}{L^2}
\sum_{\kk,\lambda}
G^{\lambda,\lambda}_{\kk,-\kk} \int_{-\infty}^{\infty} dt P(t) \langle
\tilde{R}^\lambda_\kk(t) \tilde{R}^\lambda_{-\kk}(t) \rangle 
\nonumber \\
&=& 
(-i) \frac{2 \bar{G}}{\Gamma} \frac{1}{L^2}
\sum_{\kk, \lambda} G^{\lambda,\lambda}_{\kk,-\kk} \bar{x}_{\lambda,\kk}^2 (2 n_{\lambda,\kk} + 1) .
\end{eqnarray} 
where we used $\bar{x}_{\lambda,\kk} =
1/\sqrt{\omega^\lambda_\kk/\omega_\xy}$, and $n_{\lambda,\kk}$ is the
occupation number of the mode with polarization $\lambda$, and
wave--vector  $\kk$.
Note that this is purely imaginary, thus it should not contribute to the
error, after a correction of the gate time.
\subsection{Second order in $H^\ah_\xy$} 
We find two connected terms: 
\begin{eqnarray} 
{\cal E}^{(2)} &=& 
- \frac{1}{2 L} \bar{F}^2 \sum_{\qq,\lambda} F^\lambda_\qq
F^\lambda_{-\qq} 
\nonumber \\
& & \int P(t_1) D^{\lambda,\lambda}_{\qq,-\qq}(t_1-t_2) P(t_2) d t_1 d t_2 .  
\label{error.2}
\end{eqnarray} 
We solve the integration in frequency space: 
\begin{eqnarray} 
&& \int P(t_1) P(t_2) D^{\lambda,\lambda}_{\qq,-\qq}(t_1-t_2) d t_1 d t_2 
=
\nonumber \\
&& \int \frac{d \omega}{2 \pi} P(\omega)^2 D^{\lambda,\lambda}_{\qq,-\qq}(\omega) =
\nonumber \\ 
&&
(2 n_{\lambda, \qq} + 1)
\bar{x}^2_{\lambda,\qq} \frac{(2 \Gamma)^2}{(\Gamma^2 + (\omega^\lambda_\qq)^2)^2} 
\nonumber \\
&&
- 2 i
\bar{x}^2_{\lambda,\qq} \left( \frac{\omega^\lambda_{\qq}}{\Gamma} \frac{1}{\Gamma^2 +
(\omega^\lambda_\qq)^2} + \frac{2 \omega^\lambda_\qq \Gamma}{(\Gamma^2+(\omega^\lambda_\qq)^2)^2}
\right) .  
\end{eqnarray} 
Since $(\bar{F}/\Gamma) \approx X_0/d_0$, it is
convenient to rewrite this expression in the following form.
The second disconnected contribution of order two in $H^\ah_\xy$, is:
\begin{eqnarray} 
{\cal E}^{(3)} &=& 
- \frac{1}{L^4}
\bar{G}^2 \sum_{\qq,\kk,\lambda,\lambda'} G^{\lambda, \lambda'}_{\qq,\kk}
G^{\lambda,\lambda'}_{-\qq,-\kk} 
\nonumber \\
& &
\hspace{-0.5cm}
\int \frac{d \omega d \omega'}{(2 \pi)^2}
P(\omega + \omega')^2 
D^{\lambda,\lambda}_{\qq,-\qq}(\omega) 
D^{\lambda',\lambda'}_{\kk,-\kk}(\omega') .
\end{eqnarray}
The integrations are easily solved in the complex plane by contour
integration, however, the resulting expressions are cumbersome, and
bring very little to the discussion of the results. The limit of
high temperatures, which is valid for most of the temperature range
considered in Fig. \ref{errT}, is more easily handled:
\begin{eqnarray}
{\cal E}^{(3)} &=& 
- \frac{1}{L^4}
\bar{G}^2 \hspace{-0.25cm} \sum_{\qq,\kk,\lambda,\lambda'} 
G^{\lambda, \lambda'}_{\qq,\kk}
G^{\lambda,\lambda'}_{-\qq,-\kk} 
\nonumber \\
& &
\hspace{-0.75cm}
n_{\lambda,\qq} n_{\lambda',\kk}
\left(
2 P(\omega^\lambda_\qq + \omega^{\lambda'}_\kk) + 
2 P(\omega^\lambda_\qq - \omega^{\lambda'}_\kk) \right) .
\label{error.3.highT}
\end{eqnarray}
Note that $P(\omega^\lambda_\qq - \omega^{\lambda'}_\kk)$ contain
terms that will diverge in the limit $\Gamma \rightarrow 0$, so that,
their numerical evaluation is necesary for a quantitative
understanding of the dissipation during the quantum gate.
\subsection{Third order in $H^\ah_\xy$} 
The only disconnected contribution is given by: 
\begin{eqnarray} 
{\cal E}^{(4)} = && \hspace{-0.25cm} i \frac{1}{L^4} \bar{F}^2 \bar{G}
\sum_{\stackrel{\lambda,\lambda'}{\qq,\kk}} F^\lambda_{-\qq} F^{\lambda'}_{-\kk}
G^{\lambda, \lambda'}_{\qq,\kk} \times
\nonumber \\
& & \hspace{-2.0cm}
\int \frac{d \omega d \omega'}{(2 \pi)^2}
P(\omega) P(\omega') P(\omega + \omega') 
D^{\lambda,\lambda}_{\qq,-\qq}(\omega) D^{\lambda',\lambda'}_{\kk,-\kk}(\omega') .  
\end{eqnarray}
Again, we writte here explicitly the result for high temperatures,
that is:
\begin{eqnarray} 
{\cal E}^{(4)} = && \hspace{-0.25cm} i \frac{1}{L^4} \bar{F}^2 \bar{G}
\sum_{\stackrel{\lambda,\lambda'}{\qq,\kk}} F^\lambda_{-\qq} F^{\lambda'}_{-\kk}
G^{\lambda, \lambda'}_{\qq,\kk}
n_{\lambda,\qq} n_{\lambda',\kk} \times
\nonumber \\
& & \hspace{-1.2cm}
2  P(\omega^\lambda_\qq) P(\omega^{\lambda'}_\kk) 
\left(
P(\omega^\lambda_\qq + \omega^{\lambda'}_\kk) 
+ 
P(\omega^\lambda_\qq - \omega^{\lambda'}_\kk)
\right) .
\end{eqnarray}
Where we find, again, a divergence due to resonant terms in the limit
$\Gamma \rightarrow 0$. 

\section{Short note about the Wick's theorem at finite temperature}
Since it seems that there is some confussion in the literature about the validity of
this approach, we show here that Wick's theorem can be justified 
at finite temperatures. 

Let us define the generating functional, which functional derivation
yields the time ordered correlation functions:
\begin{eqnarray}  
&& \langle {\cal T} 
\{ R_{\kk_1}(t_1) \dots R_{\kk_n}(t_n) \} \rangle  =
\nonumber \\
&& \frac{\delta}{i \delta j_{\kk_1}(t_1)} \dots
\frac{\delta}{i \delta j_{\kk_n}(t_n)} \frac{1}{{\cal Z}_0} {\cal Z} [j(x)],
\nonumber \\
&& { \cal Z } =
\tr \{ e^{-\beta H_0} {\cal T} e^{i \sum_\kk \int_{-\infty}^{\infty} j_\kk(t) X_\kk(t) dt} \}.
\label{generating.functional}
\end{eqnarray}
This could be the starting point for defining a path integral
representation, but it is much easier to calculate explicitly the
time--ordered exponential. 
First of all a small detail: it is convenient to consider $j_\kk(x)$
as belonging to the set of continuous functions that satisfy
$j_{-\kk}(x) = j_\kk^{*}(x)$, so that the exponent is
anti--hermitian. Note that the generating functional
(\ref{generating.functional}) has the same form as the evolution
operator of a forced harmonic oscillator, thus, its form can be
explicitly obtained, by following the same lines. In any case, since
the conmutators of the exponents at different times, are itself
quadratic forms of the currents $j_\kk$, it is clear that the
generating functional is gaussian. Since its second functional derivative
yields the time ordered correlation function of tow position operators, the only
choice is the following one:
\begin{eqnarray}
&& \hspace{-.5cm}
{\cal Z} [j(x)] / {\cal Z_0} = 
\\
&& 
\exp \sum_\kk \left( + \frac{1}{2} \int_{-\infty}^{\infty} j_{-\kk}(t_1)
  D_\kk(t_1-t_2) j_\kk(t_2) \ d t_1 d t_2 \right) \nonumber
\end{eqnarray}
where the kernel of the functional, $D_\kk (t_1 - t_2)$, 
is the correlation function defined in the previous sections.
Higher order functional differentiation leads immediately to Wick's theorem.

\section{Effective spin--spin interactions induced by a
  Walking--Wave}

Finally, we introduce here a scheme for inducing effective spin-spin
interactions which relies on the use of a walking wave. This is a
different approach to the one presented in \cite{spin.simulator}, which
relies on the coupling of the internal states of the ions to the
motion by means of a standing wave. The latter approach might be
difficult to implement, since one needs to fix the position of the ions relative to
the minima of the standing wave. This condition is not necesary in the
walking wave scheme, something that may be useful in quantum simulations with
ions in Paul or Penning traps.

A walking wave, like the one used by the NIST group for the
geometrical phase gate \cite{ion.qc.experiment}, leads to a
spin--motion coupling of the following form:
\begin{equation}
H_\textmd{f} = \frac{\Omega}{2
} \left( 
e^{i k (x_0 + x_j) - i \omega_L t } +  e^{- i k (x_0 + x_j) + i \omega_L t}
\right) \sigma^z_j ,
\label{walking.wave}
\end{equation}
where $\omega_L$ is the detuning of the laser beams that create the
walking wave. We consider that ${\bf x}$ is a direction transverse to
the Coulomb crystal, that is, the radial direction in the case of a
linear chain, or the axial direction in a 2D Coulomb crystal. 
If the walking wave propagates in this direction, then all the ions are at
the same position, $x_0$ with respect to the walking wave.  Note also
that we are neglecting ac Stark shifts, which can be cancelled in a
laser configuration like the one used at NIST \cite{ion.qc.experiment}. 

We consider here that the vibrational modes of the ions
are in the ``stiff'' limit, that is, the trapping potential is much
larger the Coulomb repulsion between ions. 
The validity of this limit can be quantified by means of the following quantity:
\begin{equation}
\beta_x \equiv \frac{e^2}{m d_0^3 \omega_x^2},
\label{beta.parameter}
\end{equation}
where $d_0$ is the mean distance between ions. If $\beta_x \ll 1$,
then the energy of the vibrational modes is not strongly modified by
the Coulomb interaction, and all the modes are close in
energy. Indeed, $\beta_x$ is an approximation to the dispersion in the
energies of the vibrational modes, $\omega_n$. Let us say that
$\omega_x$ is the trapping frequency. Then, all the energies lie below
$\omega_x$, and satisfy that:
\begin{equation}
\left( 1 - \beta_x \right) \omega_x \lesssim \omega_n < \omega_x .
\label{bandwidth}
\end{equation}
\begin{figure}[!]
\resizebox{3.in}{!}{
    \includegraphics{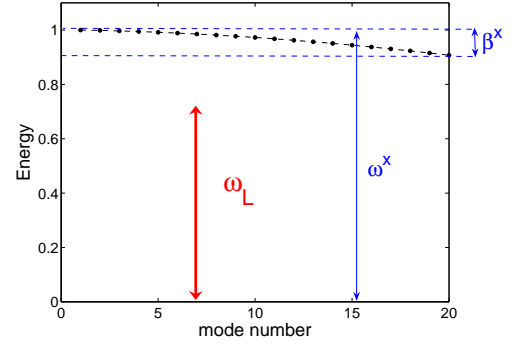}}
  \caption{Radial vibrational normal modes of a chain with 20 ions,
    and ration between radial and trapping frequencies:
    $\omega_r/\omega_z = 10$.}
\label{fig1}
\end{figure}

If we consider the Lamb--Dicke expansion of (\ref{walking.wave}), then:
\begin{eqnarray}
H_\textmd{f}(t) &\approx&
\frac{\Omega}{2} \left(e^{i k x_0 - i \omega_L t} + h.c.  \right) \sigma^z
\nonumber \\ &+&
\frac{\Omega}{2} 
\left(
i k e^{i k x_0} e^{-i \omega_L t} \sum_j x_j + h.c.
 \right) \sigma^z_j .
\label{lamb.dicke.expansion}
\end{eqnarray}

Our idea is the following: in the case $\beta_x \ll 1$, one can tune
$\omega_L$ to the red--sideband with respect to all the vibrational
frequencies in (\ref{lamb.dicke.expansion}) (see Fig. \ref{fig1}), because in this case,
ions are close to be independent. In the end, this will
lead to a coupling of the form $\sum_j x_j \sigma^z_j$, which will
induce an effective spin--spin interaction just in the same way as in
our previous proposals, with the advantage that the amplitude of the
force does not depend now on the position of the ions relative to the
phase of the walking wave.

Let us describe this situation formally. First, remember that the
position of the ions can be expressed in terms of normal modes:
\begin{equation}
x_j = \sum_n  {\cal  M}_{j,n} 
\left( \frac{\hbar}{2 m \omega_n} \right)^{1/2} \hspace{-0.2cm}
\left( a_n + a^\dagger_n \right) .
\label{normal.modes.expansion}
\end{equation}
where ${\cal M}$ are matrices that diagonalize the harmonic motion of
the ions (they are basically sines and cosines).

We want to tune $\omega_L$ to the red sideband with respect to all the
modes $\omega_n$, so that we choose 
$\omega_L < \omega_n$, and $\omega_x - \omega_n \ll \omega_n$. Under
these conditions we can keep only the red sideband terms of (\ref{lamb.dicke.expansion}):
\begin{eqnarray}
H_\textmd{f}(t) &\approx& 
\sum_n \left( g_n^* a_n e^{i \omega_L t}  
+ g_n a^\dagger_n e^{- i \omega_L t} \right) , 
\nonumber \\
g_n &=& \sum_j i k \frac{\Omega}{2} e^{i k x_0} 
\left( \frac{\hbar}{2 m \omega_n} \right)^{1/2} \! \! \! 
{\cal M}_{j,n} \ \sigma^z_j .
\label{red.sideband.2}
\end{eqnarray}

The rotation of the terms in (\ref{red.sideband.2}) can be eliminated
by considering the phonons in a rotating frame, in which their energy
is shifted by $\omega_L$. 
The whole hamiltonian in this frame reads:
\begin{eqnarray}
H &=& H_{\textmd{0}} + H_{\textmd{f}}(0) \nonumber \\
  &=& \sum_n \hbar \delta_n a^\dagger_n a_n + 
\sum_n \left( g_n^* a_n + g_n a^\dagger_n \right),
\label{hamiltonian.rotating.frame}
\end{eqnarray}
with $\delta_n = \omega_n - \omega_L$. Note that
(\ref{hamiltonian.rotating.frame}) induces a state dependent force on
all the ions simultaneously. A difference with the pushing gate approach
is that vibrational energies suffer a shift that we can controll by
tuning $\omega_L$.

To deal with this coupling we use, as usual, a canonical
transformation, to map our system to a quantum spin model:
\begin{eqnarray}
{\cal S} &=& \sum_n \left( \eta^*_n a_n - \eta_n a^\dagger_n \right)
\nonumber \\
\eta_n &\equiv& \frac{g_n}{\hbar \delta_n} .   
\label{canonical.transformation}
\end{eqnarray}
Under this transformation, the internal states of the ions follow the
dynamics of a spin--spin Hamiltonian:
\begin{equation}
H \rightarrow e^{- \cal S} H e^{\cal S} = H_0 + \frac{1}{2} 
J_{j,k} \sum_{j,k} \sigma^z_j \sigma^z_k .
\end{equation}
The effective spin--spin interaction is given by:
\begin{equation}
J_{j,k} \equiv \sum_n \frac{|F^2|}{m \omega_n \delta_n} {\cal M}_{j,n}
{\cal M}_{k,n} .
\label{effective.interaction}
\end{equation}
where is the force induced by the walking wave,  
$F = i k (\Omega/2) e^{-i k x_0}$. The spin--spin interaction
does not depend on the absolute position of the ion crystal, or
the ion plane.

Note that one important difference with our previous work is that now the
shifted frequencies appear in the denominator, thus, we can expect
that the dependence of $J_{j,k}$ on the inter--ion distance will be
different, and even suitable to be controlled by tuning $\omega_L$.

Let us now estimate the strength of $J_{j,k}$ in the simplest case. For this, let us recall that the normal vibrational modes can be written like this:
\begin{equation}
(\omega_n)^2 = (\omega_x)^2 \left( 1 + \beta_x V_n \right), 
\label{normal.modes}
\end{equation}
where $V_n$ is the contribution from the Coulomb interaction. So to say, $V_n$ contains the dispersion of the modes in Fig. \ref{fig1}. $V_n$ can be obtained from the harmonic terms of the Coulomb interaction by means of the transformation defined by the matrices $\cal M$:
\begin{equation}
\sum_n {\cal M}_{j,n} V_n {\cal M}_{k,n} = \frac{1}{|j-k|^3}.
\label{definition.Vn}
\end{equation}
On the other hand, the shifted modes fulfill that:
\begin{equation}
\delta_n = (\omega_x) \left( 1 + \beta_x V_n \right)^{1/2} - \omega_L.
\end{equation}
If these two conditions are met:
\begin{eqnarray}
\beta_x &\ll& 1, \\
\beta_x \omega_x / (\omega_x - \omega_l) &\ll& 1,
\label{condition.2}
\end{eqnarray}
then we can express $\omega_n$ and $\delta_n$ in a series in $\beta_x$, and get the following approximate relation for the effective interaction:
\begin{eqnarray}
J_{j,k} &\approx&  \beta_x \frac{1}{2} \frac{|F|^2}{m (\omega_x - \omega_L)^2} \frac{1}{|j-k|^3} =
\nonumber \\ 
&=& \beta_x \left( \frac{F x_0}{\omega_x-\omega_L} \right)^2 \frac{1}{|j-k|^3} \ \omega_x,
\label{interaction.cube}
\end{eqnarray}
where we have also used that $\omega_x - \omega_L \ll \omega_x$, and $x_0$ is the ground state size in the trapping potential. 
Note that condition $\beta_x \ll 1$ has to hold if we want to be able to resolve the red sideband simultaneously for all the vibrational modes, but condition (\ref{condition.2}), on the contrary, can be avoided, the only difference being that the effective spin-spin interaction will deviate from the $1/r^3$ decay in Eq. (\ref{interaction.cube}).

The expression (\ref{effective.interaction}) does not tell us much
about the interaction rate that can be achieved in an experiment. For
this, one has to take into account the error that is introduced in the
simulation by the coupling of the qubits (effective spins) to the
vibrational modes. 

We have already studied in detail this error (see \cite{spin.simulator}), which depends on the
displacement of the modes. The error in the quantum simulation is induced by the entanglement of the vibrational modes with the effective spins, and it can be approximated by the following quantity:
\begin{equation}
{\cal E} \approx  \eta^2 =  \frac{|F x_0|^2}{\hbar (\omega_x - \omega_L)},   
\label{eta.WWQS}
\end{equation}
which is simply an approximation to the coefficients $\eta_n$ in Eq. (\ref{canonical.transformation}). The strength of the effective interaction is $J \approx \beta_x \eta^2 \omega_x$. Thus, we find the same relation that in the case of the standing wave. 

Last, one has to take into account the effect of the first term in
Eq. (\ref{lamb.dicke.expansion}). Even in the limit $\Omega/\omega_L
\ll 1$, one could get into troubles in case that corrections of the
form $\Omega^2/\omega_L$ appear. We will see that this is not the
case, because this term is diagonal in $\sigma^z$. It can be rewritten
in the following way:
\begin{eqnarray}
H_\textmd{c} &=& \epsilon(t) \sigma^z, \nonumber \\
\epsilon(t) &=& \Omega \cos( k x_0 - \omega_L t ),
\label{carrier}
\end{eqnarray}
such that the eigenstates are the same, with time evolution given by:
\begin{eqnarray}
|\Psi (t) \rangle  &=& c_\uparrow(t) 
| \uparrow \rangle + c_\downarrow(t) | \downarrow \rangle, 
\nonumber \\
c_\uparrow(t) &=& e^{-i \int_0^t \epsilon(t') d t'}, 
\nonumber \\ 
c_\uparrow(t) &=& 
e^{-i (\Omega/\omega_L) \left( \sin(k x_0^j - \omega_L t) - \sin(k x_0^j) \right)}.
\end{eqnarray}
Thus, the effect of $H_c$ affects terms of the form $B \sigma^x$, and
operators $\sigma^\dagger$, get corrections of order $B \Omega /
\omega_L$. This means, that condition $\Omega/\omega_l \ll 1$ is
enough to guarantee that these corrections are small compared to
  the effective spin-spin Hamiltonian interaction strengths.


\begin{thebibliography}{99}
\bibitem{note.magnetic.field} We study here the limit in which the
  cyclotron and rotation frequencies satisfy the relation $\Omega_c/2 =
  \omega_{\textmd{md}}$, in which there is no magnetic field in the
  rotating frame.
\bibitem{note0} $\beta_z$ scales like $(\omega_{\xy}/\omega_z)^2 \sqrt{N}$ \cite{Dubin.99}, thus, it can be tuned by choosing the axial/radial trapping frequency ratio.
\end{thebibliography}
\end{document}